\documentclass[aps, prx, twocolumn, amsmath,amssymb, nofootinbib,floatfix,fontsize=10]{revtex4-2}
\usepackage{graphicx}
\usepackage{import}
\usepackage{hyperref}
\usepackage{color}

\begin{document}
\title{Discrete states and ballistic  interference in quantum wires\\ approaching macroscopic lengths}
\author{H. Weldeyesus}
\affiliation
{Department of Physics, University of Basel, CH-4056, Basel, Switzerland}
\author{T. Patlatiuk}
\affiliation {Department of Physics, University of Basel, CH-4056, Basel, Switzerland}
\author{C. P. Scheller}
\affiliation
{Department of Physics, University of Basel, CH-4056, Basel, Switzerland}
\author{A. Yacoby}
\affiliation
{Department of Physics, Harvard University, Cambridge, MA, 02138, USA}
\author{L. N. Pfeiffer}
\affiliation {Department of Electrical Engineering, Princeton University, Princeton, NJ, 08544, USA}
\author{K. W. West}
\affiliation
{Department of Electrical Engineering, Princeton University, Princeton, NJ, 08544, USA}
\author{Dominik M. Zumb\"uhl}
\affiliation
{Department of Physics, University of Basel, CH-4056, Basel, Switzerland}

\begin{abstract}
Increasing the size of a system showing quantum effects is a difficult task limited by decoherence, a diminishing quantum level spacing, and the effects of disorder spoiling the quantum behavior when growing in size. Systems in 1D offer very strong confinement in the transverse directions, thus generally enhancing quantum effects, but are notoriously sensitive to disorder. 
In this work, we present a system of 1D electrons exhibiting discrete quantum levels and fully ballistic coherent quantum interference with lengths of up to 18\,$\mu$m. Tunneling spectroscopy between two parallel quantum wires with a central gated segment shows intricate interference patterns exhibiting several different periods in magnetic field and density. An analysis over three different wire lengths and a comparison with single particle numerical simulations without any free parameters remarkably explains the full pattern including the observed periods. Therefore, these wires are essentially ideal 1D systems with aspect ratios approaching 1´000.
In addition, at low bias, we also observe not only the Coulomb charging energies but can clearly resolve the discrete orbital and spin states in up to 10\,$\mu$m long wires when filling 100 electrons with the center gate. This is made visible by a state-of-the-art low temperature and low noise measurement system. The spin filling sequence is completely regular, strictly alternating spin up and down, avoiding high spin states, while the peak conductance is modulated in accordance with the previously discussed interference patterns. 
These striking results show that single particle Schrödinger quantum mechanics such as ballistic quantum interference and discrete quantum states may be observed, under the right conditions, in systems of up to 18\,$\mu$m length, thus approaching macroscopic sizes. 

\end{abstract}
\maketitle
\section{Introduction}

 Quantum effects are easily observed on a microscopic scale, but tend to disappear as the system size gets larger. Mesoscopic physics \cite{Imry1997} concerns itself with objects which are large enough to not be describable by microscopic methods, but still exhibit quantum mechanical effects. Here, the size of an object is similar to various important length scales such as the coherence length, the mean free path and the wave length of the electron wavefunction  \cite{Beenakker1991, Imry1997}. Typically the size of these systems range from a few nm  up to a $\mu$m, allowing for the observation of interference phenomena and the appearance of quantized energy levels. 

Apart from size, dimensionality plays an important role in the properties of quantum systems, among which one-dimensional (1D) electronic devices are playing a key role. These host various interesting properties such as conductance quantization \cite{vanWees1988}, Van Hove singularities, Peierls instability \cite{Peierls1955} and of course the non Fermi-liquid behavior described by the Luttinger liquid formalism \cite{Haldane1981}. These systems are created by imposing a strong transverse confinement, which is already a difficult task on its own, but is further complicated by the need for extremely clean, disorder free systems, keeping intact the properties of the 1D system on its full length. In a 1D system a scattering center cannot be avoided by going around, as is possible in higher dimensions. Therefore, 1D systems are particularly susceptible to disorder, such that even single impurities or relatively moderate potential fluctuations can lead to the wire being effectively cut into two pieces, thereby limiting the achievable length of single-mode ballistic wires.

In addition, the presence of boundaries has important fundamental consequences. First, the boundaries break translational invariance and the associated momentum conservation. In addition, a discrete energy spectrum emerges due to the finite size of the system. Momentum resolved tunneling \cite{Eisenstein1991} is a very sensitive technique to study such  phenomena. The breaking of translational invariance introduces periodic oscillations in the tunneling conductance as a function of various parameters, such as magnetic field, gate voltage and dc-bias voltage. The discrete energy spectrum can also be observed directly in the quantum dot behavior of the tunneling spectroscopy, exhibiting Coulomb diamonds and orbital levels. Yet, this technique puts quite stringent requirements on the devices used, and has so far only been successfully utilized in GaAs cleaved edge overgrowth (CEO) quantum wires \cite{Auslaender2002}, GaAs two dimensional electron gases  (2DEGs) \cite{Eisenstein1991,Hayden1991},  edge states \cite{Huber2005, Patlatiuk2018, Weldeyesus2025b},  gate defined quantum wires \cite{Kardynal1996,Kardynal1997, Jompol2009, Weldeyesus2025a} and  recently in graphene quantum twisting microscopes \cite{Inbar2023}. 

Finite size effects as mentioned above have been studied both experimentally and theoretically \cite{Tserkovnyak2002,Tserkovnyak2003, Fiete2005} including the effects of inhomogeneous density \cite{Qian2008}. These works considered finite size effects appearing as a function of dc-voltage bias \cite{Tserkovnyak2002,Zuelicke2002, Boese2001, Boese2002, Tserkovnyak2003, Auslaender2005}, or were investigating the low-density (few electrons) regime \cite{ Fiete2005,Qian2008}, where interaction and localization effects are more dominant. Similar finite size effects were also observed in parallel tunneling between finite size 2DEGs \cite{Vianez2022, vianez2023}. While the disorder potential induced formation of small quantum dots ($\sim 200\,$nm) has been seen \cite{Auslaender2000, Steinberg2006}, observation of the discrete energy levels in long quantum wires was theoretically predicted \cite{Tserkovnyak2002}, but has been elusive so far presumably due to thermal smearing.  

In this work, we present two complementary aspects of ballistic and exceptionally long but still finite wires: broken momentum conservation leading to quantum interference effects and, at low bias and low temperature, discrete, quantized energy eigenstates, made observable by a state-of-the-art low temperature and low noise measurement system. These quantum effects are all appearing in the same semiconductor device of up to 18$\,\mu$m length, which is approaching the macroscopic scale -- nearly visible by the bare human eye.

Varying the density under a central  gate we observe intricate interference patterns with various periods in magnetic field and charge density. We systematically study the interference patterns for 3 different lengths of quantum wires ranging from 6$\,\mu$m up to 18$\,\mu$m and find excellent agreement with single particle simulations, explaining the rich structure in terms the wire segments interfering ballistically along the full wire up to 18$\,\mu$m length and an a striking aspect ratio of about 1'000 given transverse sub-20~nm confinement. Remarkably, no adjustable parameters are needed for the simulation, only lithographic dimensions and values extracted from the measurements themselves, like the charge density, are used. 


When reducing both the temperature and the applied dc-bias, the measured tunnel conductance drastically changes and only separated conductance peaks due to Coulomb charging  remain. We use these to study the quantized levels in the 6$\,\mu$m and 10$\,\mu$m wires and extract the charging energy, level spacing, and the g-factor for 100 transitions. Using the interference patterns, we can identify the origin of the discrete states to be the full length of the wires, not a partial segment which has been created due to disorder. Remarkably, the wires exhibit a  perfect spin filling sequence alternating strictly between up and down spins, fully avoiding larger spin ground states. This is thought to be a consequence of the 1D nature of the wires and the relatively small exchange due to screening.

\section{Tunneling spectroscopy}
\subsection{Sample description}
\begin{figure}[t!]
\centering
\includegraphics[width=8.5cm]{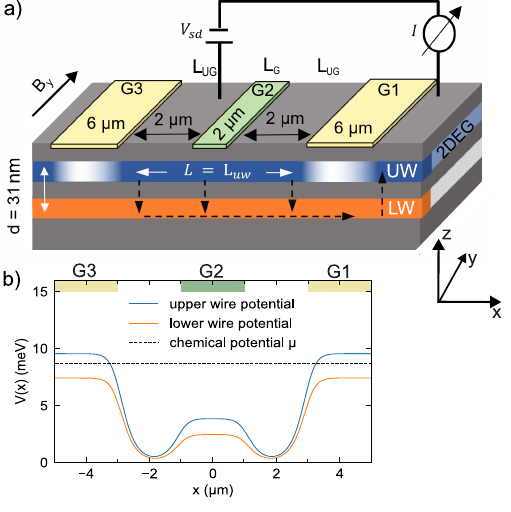}
\caption{Device sketch and potential landscape, not to scale. (a)  View onto the cleaved side of the sample. The parallel wires have a vertical center-to-center distance of $d=31$\,nm.  The shown setup corresponds to the $2|2|2$ configuration, with a 2\,$\mu$m gated section surrounded by two 2\,$\mu$m sections. The outer gates G3 and G1 define the finite-length upper wire between them, while the lower wire is not fully depleted, carrying one mode. The center gate voltage $V_{G2}$ is varied during the measurements. (b) Calculated potential profile for the upper wire (blue) and lower wire (orange). The dashed line is the chemical potential $\mu$ in the upper system. }
\label{fig:sample}
\end{figure}

The device used is a CEO double quantum wire system \cite{Pfeiffer1993} consisting of two GaAs quantum wells (QW) overgrown on one side with AlGaAs. Further details on the fabrication process can be found in the methods section. The quantum wires exist on the edge of the sample, where the upper wire (UW) is coupled to the adjacent upper 2DEG. The lower wire (LW) is only tunnel coupled to the UW / upper 2DEG while the lower 2DEG remains populated only at the edge.  Tungsten stripe gates, spanning the full width of the sample surface, are deposited on the surface of the device, 500\,nm above the upper QW. The gates have alternating widths of 2\,$\mu$m and 6\,$\mu$m and are separated by 2\,$\mu$m gaps from each other. Ohmic contacts to the QWs are made in between the gates, far away from the cleaved edge where the quantum wires are located. A schematic depiction of the sample is shown in Fig. \ref{fig:sample}(a).

A typical conductance measurement across gate G2 is shown in the inset of Fig.\ref{fig:large_scale} (a). The upper 2DEG is depleted at about -600\,mV at which point only the wire modes in the UW and LW are conducting. Decreasing the gate voltage even more to about -800\,mV depletes the UW while keeping the lower wire mode populated. The lower wire depletes at about -1'000\, mV, and no current can be passed below that gate.

\subsection{Momentum resolved tunneling spectroscopy} 
In this work we make use of the momentum-conserving tunneling between parallel quantum wires. 
When tunneling between two infinitely long 1D systems, momentum has to be conserved due to translational invariance. Since generally the density and thereby the Fermi wave number $k_{F}$ of UW and LW are different, the mismatch $\Delta k_{F}$ will suppress tunneling due to momentum conservation. Using a magnetic field $B_y$ perpendicular to the plane spanned by the two wires, a momentum boost $Q_B= e B_y d /\hbar$ is given to the tunneling electron via the Lorentz force. Here $e$ is the magnitude of the electron charge,  $d$ is the center-to-center separation between the wires, and $\hbar$ is the reduced Planck constant. Therefore, the magnetic field $B_y$ can be used to match the wave numbers of electrons tunneling between the wires. This matching will lead to a peak in the differential conductance when $k_{F}^{U}\pm eB_yd/\hbar = k_{F}^{L}$ \cite{Auslaender2002}, where $k_{F}^{U}$ and $k_{F}^{L}$ are the Fermi wave numbers in the UW and LW respectively. 

By applying a negative voltage $V_G$ to a top gate the density $n=\pi k_{F}/2$ within both wires will be reduced, which changes the magnetic field at which the momentum matching condition will be fulfilled \cite{Steinberg2006}. For a given gate voltage, and therefore density mismatch under the gate, there are two\footnote{This is for positive $B_y$. In the absence of a  finite out-of-plane magnetic field the measurements are symmetric in $B_y$} magnetic field values where the momentum matching condition is satisfied. These are given by 
\begin{equation} 	
\label{eqn:match}
B_y^{\pm}=|k_{F}^{U}\pm k_{F}^{L}|\hbar/ed .
\end{equation} 
Here, the minus sign corresponds to the co-propagating transition where the electrons retain their movement direction, while the plus sign is the counter-propagating transition, where right movers become left movers and vice versa.
When measuring the differential conductance $g(B_y,~V_{G})$ a peak appears at the position where Eq. (\ref{eqn:match}) is fulfilled. For finite length wires the differential conductance will acquire additional peaks at finite momentum mismatch, spaced at multiples of  $\Delta k =  2\pi/L$. These finite size effects strongly depend on the potential along the longitudinal direction and can therefore be used to investigate the potential landscape in the wires. 

We note that tunneling happens across the whole length of the junction. The tunnel current is proportional to the tunneling matrix element $M=\int_0^L \psi(x)\phi(x)e^{iQ_Bx} dx$ which is essentially the overlap of the wavefunctions in the x direction taking into account the momentum kick $Q_B$ caused by the magnetic field. Here it is assumed that the z and y components of this overlap are constant since these are given by the transverse confinement \cite{Patlatiuk2020}.

\subsection{Finite length tunnel junctions} 
In this section we discuss how the potential along the wire will modify the simple picture of a single conductance peak described above.         

The devices consist of gated and ungated regions, defining wire segments with varying densities, see Fig. \ref{fig:sample} (a). 
Using two outer gates, the barrier gates, a finite-length region is defined in the upper system, shown in Fig. \ref{fig:sample} (a) for the shortest wire length. While the transverse confinement is defined by the growth (about 20\,nm in each direction), we can choose the length of the wire by selecting different gates, and can reach aspect ratios of up to 1,000 for the cases shown in this work.

The voltages on the barrier gates are chosen such that the 2DEG and UW modes are depleted, while a single mode remains in the LW. Using a third gate positioned between the barrier gates, the potential in the center of the wire can be controlled and is used to create a potential bump \cite{Steinberg2006}. In this way, a wire of total length $L=A+B+C$ is created, consisting of 3 regions $A|B|C$, where $A$ and $C$ are the lengths of the ungated sections and $B$ is the length of the gated region. Here, we focus on the symmetric case where $A=C$. More general cases are discussed in Appendix \ref{asym}.    

The example of the shortest wire length of 6\,$\mu$m is shown in Fig. \ref{fig:sample}(a), where gates G1 and G3 are the barrier gates, and G2 is the central gate. Different length can be created using additional gates (not shown). The lithographic edge of the barrier gates is at $\pm$3\,$\mu$m where the UW potential crosses the Fermi level, making it a finite size system, while the LW remains continuous over several mm and is therefore considered infinitely long, Fig. \ref{fig:sample}(b). While the density in the central region is reduced with its gate, the neighboring sections are almost unaffected and remain close to the original density. Due to screening, the effects of the top-gates are more pronounced in the upper system, leading to depletion already at less negative voltages than the lower system.    

The tunneling transconductance as a function of center-gate voltage $V_{G}$ and the magnetic field $B_y$ is displayed in Figure \ref{fig:large_scale}(a), measured by modulating the gate voltage with a small ($\sim5$\,mV) ac-voltage at a few Hz. The modulated current is then measured by a lock-in amplifier, which gives a signal corresponding to $\partial I/\partial V_G$. We note that the transconductance, as measured here in response to a small modulation on the center gate, is highly sensitive to gate-voltage dependent features and mostly eliminates a gate-voltage independent background. The most striking feature is the almost parabolic-looking resonance curve, see blue dashed curve, where Eq. (\ref{eqn:match}) is satisfied underneath the gate \cite{Steinberg2006}. Upon closer examination, we observe a range of oscillating patterns, which are highlighted in Fig.\ref{fig:large_scale} (b). These correspond to finite-size effects \cite{Tserkovnyak2002, Tserkovnyak2003, Steinberg2006} arising from different regions in the tunnel junction and will be discussed in detail in the following section.        

Due to the finite length of the UW, the electronic spectrum along the longitudinal direction ($\mathbf{\hat{x}}$) is expected to be discrete. Whether the discrete spectrum can be observed, depends on the experimental conditions, in particular temperature $T$ and the applied dc voltage $V_{SD}$, both of which can be experimentally controlled. To observe the discrete energy levels, the corresponding energy scales should be smaller than the addition energy in the wire. First, we will discuss the case where these conditions are not met, and the system is in the continuum regime.

\section{Quantum interference in the continuum regime}
At large enough dc-bias $eV_{SD}\gg \Delta$ or  large enough temperature $k_BT\gg \Delta$, the electronic structure can be viewed as a continuum, with $\Delta$ being the level spacing of the states in the first UW-mode. The following measurements were taken at $V_{SD}=100\,\mu \mathrm{V}$. This is already bigger than the level spacing, even for the shortest wire length of 6\,$\mu m$, which can be readily estimated for an infinitely deep potential well, which yields a level spacing of 76\,$\mu e\mathrm{V}$. 
\begin{figure}[htb!]
\centering
\includegraphics[width=8.6cm]{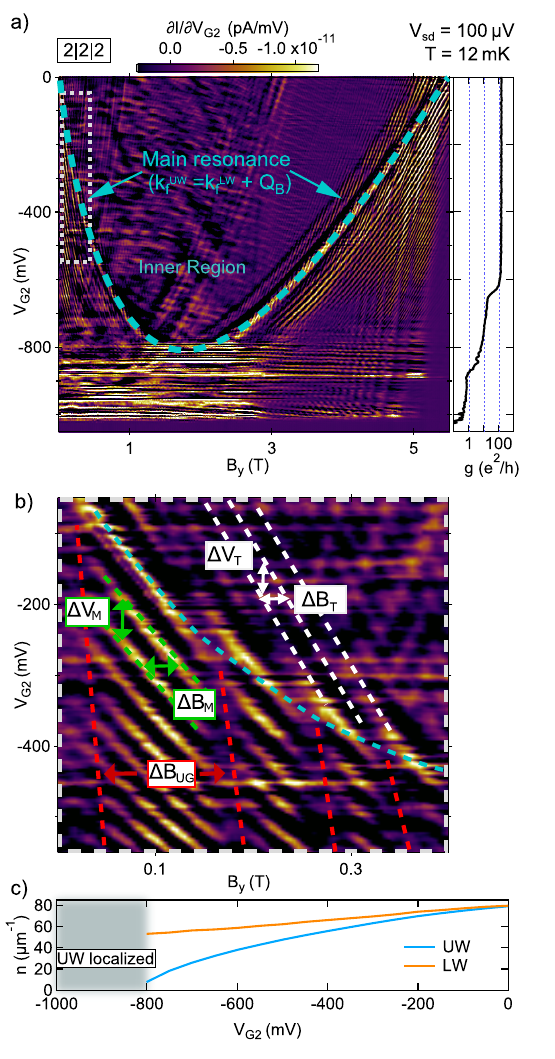}
\caption{(a) Transconductance $\partial I /\partial V_{G2}$ of $2|2|2$ junction over the full range of gate voltage and magnetic field measured at 100\,$\mu$V dc-bias. The main tunneling resonance between the lowest UW and LW modes is marked with a blue dashed curve. The inset shows $g$ vs $V_{G2}$ with all other gates deactivated, showing 2DEG and eventually wire mode depletion. The white box is magnified in (b), showing finite size effects arising from the gated section (green), ungated section (red) and the total length (white). The same color scale as in (a) is used. The extracted values are: $\Delta V_M = 51$\,mV, $\Delta V_t = 46.2$\,mV, $\Delta B_M = 38$\,mT, $\Delta B_{UG} = 99$\,mT, $\Delta B_T = 25$\,mT. (c) UW and LW density extracted from the main resonance in (a) using Eq.~(\ref{eqn:match}). Due to screening the UW depletes before the LW.}
\label{fig:large_scale}
\end{figure}

We begin our discussion with a specific example of a 6\,$\mu$m long wire where we highlight the occurring features and their possible origins, before we move to a comparison of different length wires.  The magnified view of the tunnel conductance of the $2|2|2$ configuration in Fig. \ref{fig:large_scale}(b) shows the four associated finite size quantum interference effects.
The main resonance (blue dashed curve) is broken up into segments. For gate voltages more negative than the one at which the main resonance appears, we observe replicas of the main resonance (green dashed lines). Both the main resonance and the replicas get interrupted by almost vertical stripes of vanishing oscillations (red dashed lines). For gate voltages less negative than, i.e. above the main resonance, we observe a different set of resonances (white dashed lines) which have a different slope than the main resonance replicas. The stripes of vanishing oscillations terminate at the main resonance and do not interrupt the "white" resonances.

Features with opposite slopes (e.g. rising in $V_{G2}$ for increasing $B_y$) to the ones described above can be seen in the top right corner of Fig. \ref{fig:large_scale}(b) and are understood as tunneling from the 2DEG to the second lower wire mode and are not discussed further. We proceed to provide explanations for these main observations as described above. Here, we find  that all oscillations are caused by the finite size sections appearing in the wire and the corresponding period is related to that length $L_i$ by $\Delta k = 2\pi n/L_i ,n\in \mathbb{N}$.

\textbf{Main resonance replicas (green dashes):}    
When reducing the gate voltage of the center gate, the density in both UW and LW is also reduced. The local Fermi wave number of electrons under the gate is reduced according to $k_{F}=\pi n/2$. This means the momentum mismatch between the UW and LW under the gate is given by 
\begin{equation}
\Delta k_{F}= \pi(n_{U}(V_{G})-n_{L}(V_{G}))/2+eB_yd/\hbar,
\label{eq:dkf}
\end{equation}
where $n_{U}(V_{G})$ and $n_{L}(V_{G})$ are the UW and LW density under the gate,respectively (see Fig. \ref{fig:large_scale} (c)),  extracted from the main resonance and  Eq.~(\ref{eqn:match}). The condition for a replica peak to arise is that the momentum mismatch changes by a multiple of $2\pi/L$ such that:    
\begin{equation}
\Delta k^{i+1}_{F}-\Delta k^{i}_{F}= 2\pi n/L_M,
\end{equation}      
where $\Delta k^{i}_{F}$ is the momentum mismatch for the {$i$-th} replica. This allows us to extract the length associated with that oscillation period by utilizing Eq. (\ref{eq:dkf}) either at constant $V_G$ or constant $B_y$. The corresponding period in gate voltage and magnetic field are $\Delta V_M$ and $\Delta B_M$, which interestingly, seem to yield two different length $L_{M_V}$ and $L_{M_B}$ , respectively. At constant $V_G$ this gives a length of $\sim 3.5$ $\pm$ 0.2\,$\mu$m while at constant $B_y$ a length of  2.3$\pm$0.4\,$\mu$m results. The gate dependent oscillation yield a length in good agreement with the lithographic gate width, the magnetic field period does not. In fact it is close to the value expected for the sum of gate width and ungated section width (i.e A+B or B+C). This finding is confirmed by the other configurations and simulations where the lengths of the gated and ungated section are varied. In all configurations, $L_{M_V}$ is close to the lithographic gate length, and $L_{M_B}$ is similar to the sum of the two sections. 

Generally, we find the oscillation frequency in gate voltage to increase slightly at more negative voltages because the gating of the UW becomes more effective due to less screening at lower densities. This can be seen clearly in Fig. \ref{fig:large_scale}(c) where the $UW$ density reduces faster as $V_{G2}$ is decreased, while the density in the $LW$ decreases linearly. This means the rate at which the density-difference and therefore $\Delta k_{F}$ grows increases as $V_{G2}$ is made more negative. When displaying the data as a function of the density difference $\Delta n (V_G) = n_{U}(V_{G})-n_{L}(V_{G})$ the spacing between the resonances becomes equidistant (see Appendix \ref{add_data}).

\textbf{Stripes of vanishing oscillations (red dashes):}
The regions on either side of the central gate are only weakly influenced by changing the gate voltage. While there is a slight slope on the stripes in $V_G$ when the ungated section is 2\,$\mu$m,  almost no slope is seen if it is 6\,$\mu$m long, as can be seen for the $6|6|6$ configuration further down. Therefore, they only contribute a gate-independent modulation of the overall oscillation pattern\footnote{It is important to note, that the stripes of vanishing oscillations we discuss here are different from the ones discussed in \cite{Tserkovnyak2002, Tserkovnyak2003}, which have been attributed to the presence of two velocities due to spin-charge separation. These were observed in the $V_{SD}$ and $B_y$ plane, contrary to the measurements here in the $V_G$ and $B_y$ plane at a low bias of 100\,$\mu$V. }    for gate voltages more negative than the main resonance. This period is given by $\Delta B_{UG} = 2\pi\hbar/edL_{UG}$, with $L_{UG}$ being the length of the ungated section of the UW. In the $2|2|2$ case where the ungated section is of lithographic length 2\,$\mu$m,  we extract $L_{UG} = 1.5$\,$\mu$m.

\textbf{Inner region oscillations (white dashes):}
At fixed gate voltage, there is a fixed mismatch between the Fermi wave vectors in the UW and LW. For magnetic fields in between the matching points, oscillations appear with a period given by $\Delta B_T = 2\pi\hbar/edL_{T}$, where $L_{T}$ is the total length of the UW. We extract a length of $L_{T}=5.1$\,$\mu$m, which is slightly shorter than the lithographic length of 6\,$\mu$m. The magnetic field period is, except for the experimental data of the $6|6|6$ configuration, independent of gate voltage, and the details of the origins of this period are discussed further below. For the $6|6|6$ configuration we find a range of $L_T$ between 13\,$\mu$m and 17\,$\mu$m, with the larger values appearing at at less negative gate voltages. We therefore assume this dependence of $L_T$ on disorder to be associated with the influence of disorder as density is lowered sufficiently. The gate voltage periods of the inner region oscillations are identical with those of the main replicas, and therefore reflect the length of the gated section.  

\subsection{Period scaling}    
\begin{figure*}[htb!]
\centering
\includegraphics[width=15.5cm]{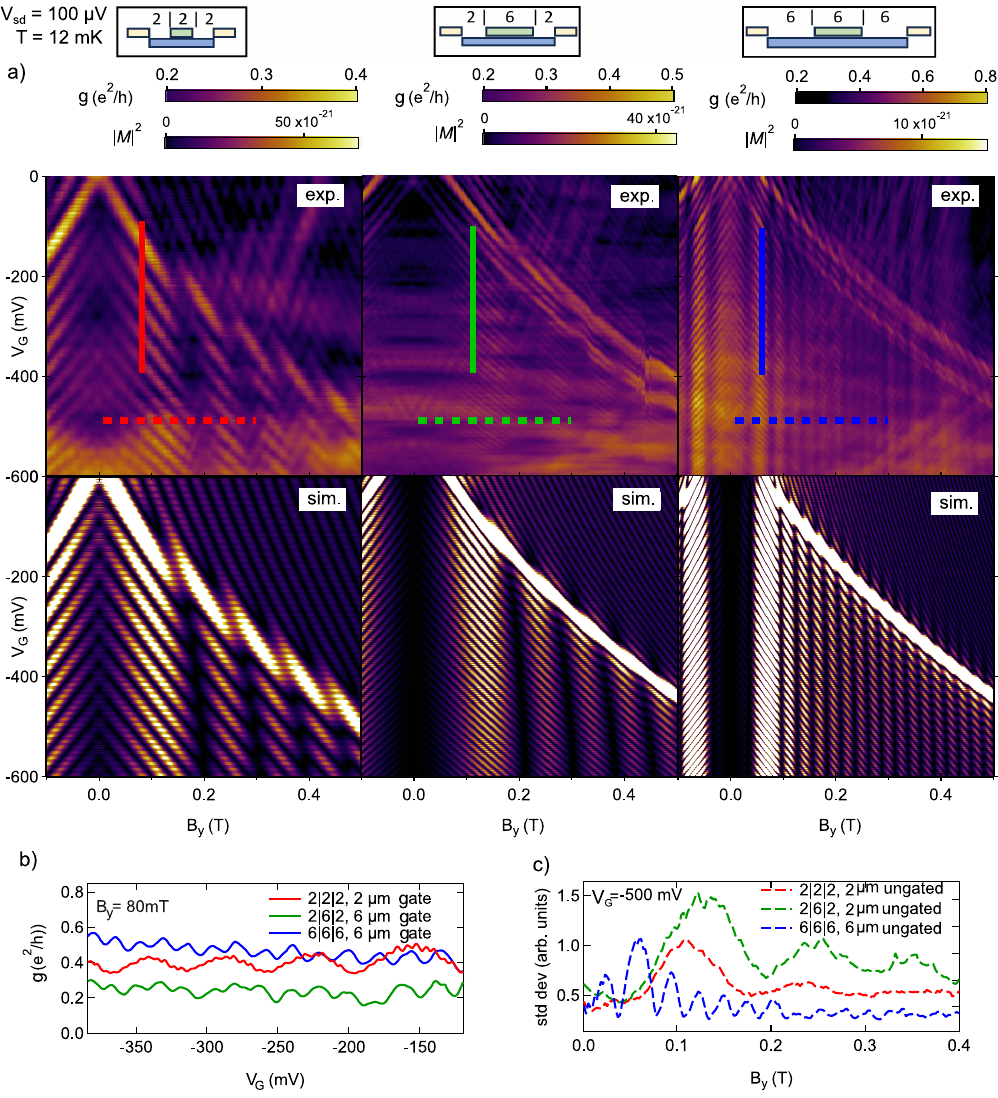}
\caption{Comparison of experiment (upper panel) and simulations (lower panel) for different wire and gate length (a) left to right: $2|2|2$; $2|6|2$; $6|6|6$. (b) vertical line cuts (offset for clarity) along solid lines in (a)  oscillation frequency reduces with shorter gate length. (c) average standard deviation along dashed lines in (a) vertical line spacing depends on length of the ungated region. }
\label{fig:meas_sim}
\end{figure*}

To further support these findings, we repeated the same analysis for the $2|6|2$ and $6|6|6$ configurations by using different gates on the same sample and observe a scaling of the periods with the various lengths. To confirm our results we perform numerical simulations (described in Appendix \ref{simulation}) to obtain the the tunneling matrix element $M \propto g$ for the three configurations  $2|2|2$, $2|6|2$ and $6|6|6$. The results are shown Fig. \ref{fig:meas_sim} (a), and display all previously discussed features, such as the correct position of the main resonance, the appearance of the main resonance replicas, and the almost vertical stripes of suppressed interference due to the ungated section. Also, the inner region oscillations are correctly reproduced. Most notably, this is achieved without the need for any tunable parameters. All relevant quantities are either given from the lithographic dimensions, or extracted from the tunneling spectroscopy measurements directly, such as the density of the UW and LW modes. Thus, the agreement, over the three different length configurations, is quite striking, and demonstrates a very thorough understanding of the double wire system. The extracted lengths from measurements and simulations all agree well with the lithographic dimensions of the device and with each other and are summarized in Table  (\ref{tab:lengths}). 

We show the scaling of the periods with gate length and ungated section length in Fig. \ref{fig:meas_sim} (b) and (c), displaying a clear tripling in the frequency when the corresponding length is increased from 2\,$\mu$m to 6\,$\mu$m, confirming the assumption of $\Delta k_F = 2\pi/L_i$. 

Especially for the $2|2|2$ and $2|6|2$ we find excellent agreement between simulation and experiment for all extracted values. The fact that we extract a similar $L_{UG}$ in both cases, where the lithographic length is the same, is noteworthy.  Also, the total length $L_T$ is reproduced well, and in agreement with the value extracted from the simulations. In general, we observe the extracted length for the gated section to always be bigger than the lithographic length, while the ungated section is always smaller.  This can be explained by a gating effect outside the lithographic gate dimension, making the gated region $L_G$, and the regions under the  two barrier gates effectively bigger, at the cost of the ungated sections. 

When comparing the the $\Delta B_{M_G}$ periods of the $2|6|2$ and $6|6|6$ configuration we find  the extracted lengths to differ even though the lithographic dimensions of the gated sections are the same. The values are in good agreement with the length of the sum of gated and ungated section which is 8\,$\mu$m (7.8\,$\mu$m extracted) for the $2|6|2$ configuration and  12\,$\mu$m (11\,$\mu$m extracted) for the $6|6|6$ configuration.

For interference over the full length of the combined segments,  there is excellent agreement between simulation and experiment for the two shorter configurations, see the $L_T$ entry in the table. For the long $6|6|6$ configuration, the extracted length depends on gate voltage. At more negative voltages, where the density is lower and correspondingly also a higher sensitivity to disorder appears, the extracted lengths are a few microns below the full length. This is presumably due to disorder making the effective length a bit shorter. Nevertheless, at voltages close to zero, corresponding to high carrier densities well above the disorder potential, the oscillation period corresponds to the full expected length of $\sim$17\,$\mu$m, indicating full ballistic interference over the entire length of the wire. 


The previously mentioned resonances due to the presence of higher modes (running diagonally toward the upper right in Fig. \ref{fig:large_scale} (a)) are not reproduced in the simulations because only  the lowest modes were considered. 
The fact that the main resonance for the $2|6|2$ and $6|6|6$ configurations (the two arrangements of largest length) appears to be made up of two resonances, which also is not reproduced by the simulation. A possible explanation could be given by the presence of a density variation in the UW underneath the gate. The needed difference in density to reproduce the observed splitting of $\Delta B \sim 100$\,mT  can be estimated as $\Delta n= 2\Delta B e d / \pi \hbar = 3 \,\mu \mathrm{m}^{-1}$, while the total density without the effect of the gate is $\sim 80$\,$\mu$m$^{-1}$. A simulation showing this double peak structure, and the assumed potential profile, is shown in Appendix \ref{doubling}.

\begin{table}
\centering
\begin{ruledtabular} 
\begin{tabular}{p{17mm}p{7mm}|ll|ll|ll} 
$L_{UG}|L_G|L_{UG}$ & $[\mu \mathrm{m}]$ & \multicolumn{2}{c|}{$2|2|2$} & \multicolumn{2}{c|}{$2|6|2$} & \multicolumn{2}{c}{$6|6|6$} \\ 
\hline
&  & Exp. & Sim. & Exp. & Sim. & Exp. & Sim.\\
\hline\hline
$\Delta B_M $ $(\pm 2)$ & [mT] & 38 &38& 17.3 &17& 12.4 &11\\ 
$\Delta B_{UG}$ $(\pm 5)$ & [mT]& 99 &97 &94 &87& 26.2 &26.7\\ 
$\Delta B_T$ $(\pm 2)$& [mT]& 25.8 &26& 13.5 & 14& 8-11 &7.6\\ 
\hline 
$L_{M_B} $ & $[\mu \mathrm{m}]$& 3.5 &3.5& 7.8 &7.8& 11 &12\\ 
$L_{UG}$  & $[\mu \mathrm{m}]$ & 1.3 &1.4& 1.4 &1.5& 5.1 &4.9\\ 
$L_T$ & $[\mu \mathrm{m}]$& 5.1 &5.1& 9.8 &9.5& 13-17 &16.5\\ 
$L_{M_B}+L_{UG}$ & $[\mu \mathrm{m}]$& 4.8 &4.9& 9.2 &9.3& 16.1 &15.9\\ 
$L_{litho}$ & $[\mu \mathrm{m}]$& \multicolumn{2}{c|}{6} & \multicolumn{2}{c|}{10}&\multicolumn{2}{c}{18}\\ 
\end{tabular}
\end{ruledtabular}
\caption{Extracted lengths from finite size effects. For each entry we compare the numbers acquired from experiment and simulation. $L_{M_B}$ is the  length of the sum of gated and ungated section (see text). $L_{UG}$ is the length of the ungated section alone, extracted from the period of the stripes of vanishing oscillations. $L_T$ is the full length of the wire extracted from the inner region oscillations. $L_{litho}$ is the lithographic full length of the wires. The errors in $L$ scale as $1/\Delta B$ and are therefore more pronounced for small $\Delta B$ (i.e. long length)}
\label{tab:lengths}
\end{table}

\begin{figure*}[htb!]
\centering
\includegraphics{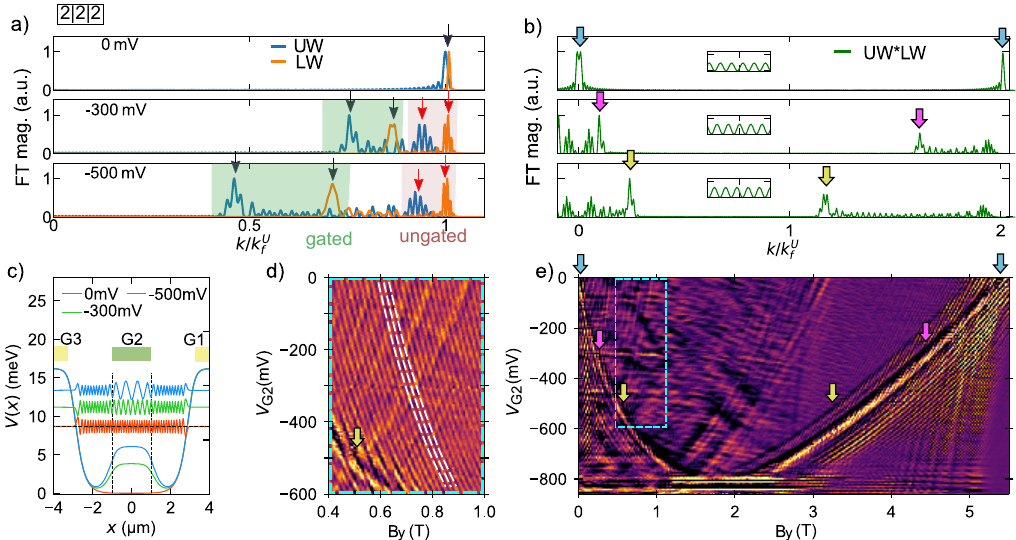}
\caption{Spectral composition: (a) Fourier transforms of UW (blue) and LW (orange) wave functions. The wave vectors under the gate G2 are highlighted with black arrows (green shaded area), while the red arrows indicate the wave vectors in the ungated regions (red shaded area). (b) Fourier transform of the product of UW and LW wave functions. The main resonance peaks are highlighted with colored arrows. The inset shows a magnified view (x70) of oscillations in the region between the marked peaks. (c) Potential shape in the UW for three gate voltages (0\,mV, -300\,mV, and -500\,mV). The corresponding wave functions are shown (at much lower density for better visualization), offset for clarity, at the chemical potential $\mu$ (horizontal dashed line). Vertical dashed lines show the lithographic extent of the gate G2. (d) Derivative with respect to magnetic field of conductance inside the blue dashed box marked in (e), highlighting the regularity of the inner region oscillations (diagonal white dashed curves), and the independence of the period on magnetic field and gate voltage. Features from lower left to upper right correspond to tunneling from higher modes. (e) correspondence between features in the tunnel map and peaks in the product Fourier spectrum. Data shown is the same as in Fig. \ref{fig:large_scale} (a).}
\label{fig:fft}
\end{figure*}

\subsection{Spectral analysis of wave functions}
Figure \ref{fig:fft} (a) shows the Fourier transform (FT) magnitude of $UW$ and $LW$ wave functions for three different gate voltages. At $V_G=0$ the spectrum is strongly peaked at a single wave number $k_F$ (black arrow in the first panel) with a width on the order of  $2\pi/L$ for the UW. The lower wire is open and thus shows a sharper Fourier peak. As expected, the spectrum contains no weight at wave numbers larger than the Fermi wave number, since higher energy states are not occupied. Due to the soft longitudinal confinement $V(x)$, wave numbers $k<k_{F}^{U}$ are added to the spectrum, but for $V_G=0$ most of the spectral weight is concentrated around $k_F$. If one now increases the gate voltage $V_{G}$, the local density under the gate, and therefore the local Fermi wave number $k_{F}(x)$, is reduced. This introduces a second peak in the FT at smaller wave numbers. These peaks are marked with black arrows in the second and third panels. Close to  $k_F^U$ a peak remains in both UW and LW spectra, corresponding to the dominant wave number in the ungated section. 
This peak acquires a substructure and broadens, while also shifting to lower values, because the bottom of the confinement potential is slightly pulled up. This effect is more pronounced for the upper wire due to more effective screening of the gate, as seen from the lower wire. As discussed earlier, this leads to the slight slope of the stripes of vanishing oscillations seen in the $2|2|2$ configuration in Fig. \ref{fig:meas_sim} (a).  

When forming the product of the two wave functions, as required in the calculation of the tunneling matrix element, the FT of the product will contain wave vectors up to $k=k_{F}^{U}+k_{F}^{L}$. This can be easily understood for an infinitely long wire with homogenous density, where the wave functions are just plane waves $\sin(kx)$. The matrix element then becomes,
\begin{align}
M(Q_b) & =  \int_{-\infty}^\infty  \sin(k_1x)\sin(k_2x)e^{iQ_bx}dx \nonumber \\
& =  \delta(-Q_b+k_1-k_2) + \delta(Q_b+k_1-k_2)      \nonumber         \\
&  +\delta(Q_b-k_1-k_2) +\delta(-Q_b-k_1-k_2).  \label{eq:delta}
\end{align}
This essentially reproduces the behavior of Eq. \ref{eqn:match} for the infinitely long wire. In the finite length limit, other frequency components will play a role as well, which leads to a more complicated picture. 

Figure \ref{fig:fft}(b) shows the FT magnitude of the product of the UW and LW wave functions for the same gate voltage as in Fig. \ref{fig:fft}(a). The peaks from the gated section marked with black arrows in (a) lead to the difference and sum peaks marked with colored arrows in (e) when UW and LW wave functions are multiplied. The relationship between the FT of wave function products and the measurements becomes clear in Fig. \ref{fig:fft}(e). The arrows mark the positions of the main replica and separate the two regions of the FT spectrum into the inner region enclosed by the two peaks (inside the parabola), and the outer region. As discussed earlier, all the oscillations related to density inhomogeneities are contained in the outer region, while the inner region only displays oscillations related to the full length of the wire.

\begin{figure}[tb!]
\centering
\includegraphics[width=8.6cm]{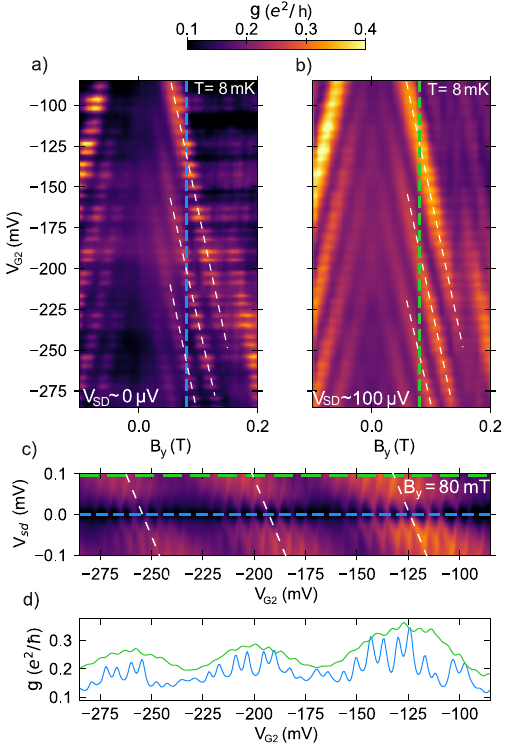}
\caption{Crossover between discrete and continuous regime for the $2|2|2$ configuration. (a)/(b) Comparison between finite size effects at $V_{Sd} \sim 0\,\mu \mathrm{V}$  and $V_{Sd} \sim 100\,\mu \mathrm{V}$.  (c) Coulomb diamonds at B\,=\,80\,mT. (d) Horizontal line cuts of (c) along the green and blue dashed lines. }
\label{fig:lowbias}
\end{figure}
Though it looks like the inner region is empty, the zoomed inset in Fig. \ref{fig:fft}(d) shows that there are small magnitude oscillations throughout the inner region. The oscillations arising here are essentially from the FT of a finite size windowing function given by the length of the upper wire. To illustrate this effect we assume a plane wave function in the upper system, without boundaries. The FT is defined with the integration boundaries ranging from $-\infty$ to $\infty$, so for the integral to be bound by the length of the wire one can include a windowing function $w(x)=\Theta (x)-\Theta (x-L)$, where $\Theta$ is the Heaviside function. $M(Q_b)$ then becomes:
\begin{equation}
M(Q_b)=\int_{-\infty}^{\infty}~\psi(x) e^{-iQ_bx}\varphi(x) w(x)dx
\end{equation}
In this sense, the wave functions $\psi(x) , \varphi(x)$ correspond to the wave functions of infinitely long wires. The finite length UW wavefunction is then given by $\psi(x) w(x)$. The FT of w(x) is a $\widetilde w(k)=L\cdot\mathrm{sinc}(kL/2)$. As a result of the convolution theorem, $\widetilde w(k)$ will be placed at the positions of the $\delta$-functions in Eq. \ref{eq:delta}, where the main peak of the sinc function corresponds to the main resonance and the side lobes to both side are  the inner region oscillations.

\section{Quantized regime}

The previously discussed oscillations in conductance can be understood completely within a continuum model, where the discreteness of energy states does not have to be considered, since a large dc-bias exceeding temperature and the level spacing was applied. When the applied dc-bias voltage is smaller than the level spacing of the UW, the assumption of a continuum of electronic states is not valid anymore, and the discrete nature of the UW states becomes relevant. In the following, we will discuss the crossover from the continuum to the discrete regime for the $2|2|2$ configuration.
\begin{figure*}[htb!]
\centering
\includegraphics[width=0.95\linewidth]{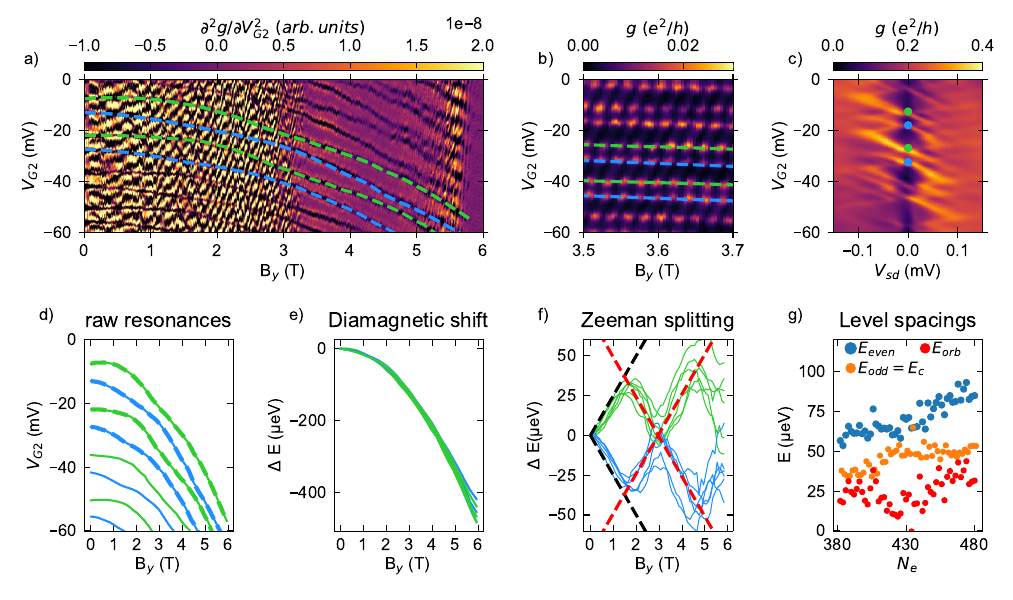}
\caption{ Energy level spectroscopy of a 6\,$\mu$m ($2|2|2$) long wire. (a) conductance map showing the first few Coulomb peaks. For magnetic fields below 3\,T additional peaks from other modes reduce visibility. The second derivative with respect to gate voltage is shown for clarity. Four resonances are marked with alternating green and blue dashed curves. (b) Zoom in into the dashed white region in (a). Dashed curves are the same as in (a). Each Coulomb peak shows periodic modulation of conductance as $B_y$ is varied, with $\Delta B = 22$\,mT. (c) Coulomb diamonds taken at $B_y=0$\,T of the first few resonances used to extract the lever arm $\alpha=9.3\,\mu$eV/mV. Blue and green dots mark the corresponding resonances in (a) and (b). d) The extracted resonances from (a).  (e) The extracted diamagnetic shift for the first 12 resonances. (f)  Extracted  Zeeman energy for the first 12 resonances. Dashed black lines correspond to the bulk GaAs g-factor of  $g=-0.44$. Red dashed lines mark the crossover from higher/lower orbitals that are lowered/lifted by the Zeeman energy. (g) Dependence of even $E_{even} (E_c)$, odd $E_{odd}$, and single particle level spacing $\Delta=E_{odd}-E_{even}$ on occupation number.}
\label{fig:6um_zb}
\end{figure*}
Figure \ref{fig:lowbias}(a) and (b) show a comparison between conductance maps measured with and without an applied dc-bias voltage. The previously discussed oscillation pattern is still visible for zero bias, but is now broken up into discrete peaks that align with the original main resonance replicas. Upon closer examination, one finds that the peaks are bunching up in pairs along the gate voltage direction, which is visible in Fig. \ref{fig:lowbias} (a) and the line cut in (d). This suggests the presence of two energy scales of similar size. 

As a function of voltage bias, a set of clear Coulomb diamonds is seen, shown in Fig. \ref{fig:lowbias}(c) at the magnetic field indicated by the blue dashed line in \ref{fig:lowbias}(a). This suggests that a quantum dot is formed here in the upper wire, associated with a Coulomb charging energy $E_c$ and a orbital level spacing $\Delta$. Due to the spin degeneracy of the orbital levels, the peak spacing depends on whether subsequent peaks belong to the same or different orbitals. In the same orbital, the addition energy is given only by the Coulomb charging energy $E_c$, while adding an electron to the next orbital results in an addition energy of $E_c + \Delta$.      
The three bright lobes that can be seen in Fig. \ref{fig:lowbias}(c) (white dashed lines) correspond to the three main resonance replicas that are crossed by the blue dashed line in Fig. \ref{fig:lowbias}(a). Their bias dependence is another manifestation of the finite size effects and follows the period $\Delta V_{SD}= \pi^{2}\hbar^{2}n/e m^{*}L_G$ \cite{Tserkovnyak2003} which corresponds to a length of  $L_G=1.9$\,$\mu$m.

Quantum dot formation in CEO wires was previously observed  at gate voltages close to the depletion of the wire \cite{Auslaender2000,Steinberg2006}, induced by the disorder potential on a length of a few 100\,nm. In contrast, here, we observe charging physics starting from the largest densities in the ballistic regime, where the size of the quantum dot is given by the full length of the wire as defined by the gates, corresponding to 6\,$\mu$m. Data from a 10\,$\mu m$ junction shows very similar behavior, see Appendix \ref{10mu}.  
Figure \ref{fig:6um_zb}(a) shows an extended magnetic field range up to 6\,T. The second derivative with respect to gate voltage is shown to highlight the Coulomb peaks. The resonance positions are extracted for clarity in Fig. \ref{fig:6um_zb}(d). Between 0\,T and 3\,T, closely packed upward moving resonances can be observed. Their origin is unclear, but we hypothesize that they might originate from the finite size states of the 2DEG, which is also confined to the same length of 6\,$\mu$m and therefore should have a similar discrete spectrum of subbands in the $\hat x$-direction as the wire mode. Alternatively, they could also be caused by the presence of an additional UW mode.  Above 3\,T these states vanish (see the transition from yellow to purple in Fig. \ref{fig:6um_zb}(a)) because the momentum shift $Q_B$ exceeds the counter-propagating transition of the modes with the LW mode. Above 3\,T, we are left with the Coulomb peaks of the first UW mode only (purple region Fig. \ref{fig:6um_zb}(a)), which are now easy to follow even into the low field region.

The resonances stemming from transitions of the UW states are overlaid by green and blue dashed curves for two sets of Coulomb peaks in Fig. \ref{fig:6um_zb}(a). Further evidence for a long quantum wire is given by the conductance modulations on the Coulomb peaks themselves, as the magnetic field is varied. As previously discussed, the finite size effects in the inner region lead to oscillations where the period only depends on the full length of the UW. In a strongly magnified window in magnetic field, see Fig. \ref{fig:6um_zb}(b), we observe a discrete set of conductance peaks which together make up the larger data set when zoomed out. The period of this modulation is $\Delta B=21$\,mT which corresponds to a length of $L_{T}=5.3\,{\mu m}$, indicating that the dot indeed extends over both gated and ungated segments. Thus, the quantum dot regime here is combined and superimposed with the coherent interference physics of the momentum resolved tunneling. This is imprinting 
discrete Coulomb peaks on top of the intricate coherent interference patterns as described in the previous section.  

Further, when looking at the bias voltage dependence, we observe clear Coulomb diamonds, see Fig. \ref{fig:6um_zb}(c),  allowing us to extract the quantum dot parameters such as the lever-arm of the gate, and the capacitances \cite{Hanson2007}. The total capacitance $C_{tot}=3.5$\,fF can be calculated from the charging energy $E_{c} =e^{2}/C_{tot}$. The main contribution to this capacitance is the drain capacitance, as extracted from the slopes of the Coulomb diamonds. Here, this is the capacitance between the UW and LW. This can be estimated as $C_{UL}=L / (\textrm{log}(1+(D/d)^{2})/4\pi\epsilon)$ \cite{Auslaender2005}, where $D = 500$\,nm is the distance to the top gate, $L$ is the length of the wires, and $d = 31$\,nm is the inter wire separation. This agrees well with the measured value. The lever-arm of the gate $\alpha = C_{g}/C_{tot}$ is 9.6$\times10^{-3}$ and gives a gate capacitance of $C_{g} = 33.6\,\textrm{aF}$. This also agrees very well with earlier measurements on similar devices \cite{DePicciotto2008}, even though they were acquired through different methods. These capacitances suggest that the Coulomb diamonds are caused by charging of the full wire.

From the Coulomb peaks at $B_y=0$, we extract the  addition energies for the first 100 resonances, a bit more than half of the electrons underneath the gate (n $\sim$ 80\,m$^{-1}$). The charging energy can be extracted from the size of the smaller (odd) diamonds, while the difference of big and small diamonds gives the orbital energy splitting $\Delta$. All three energies are displayed in Fig. \ref{fig:6um_zb}(g) revealing the clear even-odd effect discussed before. Similar behavior was observed in carbon nanotubes 
\cite{Liang2002,Cobden2002,Moriyama,Annabi2024}, but with a period of 4 reflecting the additional valley degeneracy. The general trend is a reduction of those energies, as more electrons are removed from the wire. The orbital energies appear to be increasing for higher occupation numbers $N$, see the red  markers \ref{fig:6um_zb}(g). For a potential well with infinitely steep walls, the increase  of the orbital energy spacing would be linear, while for a parabolic potential it would be constant. For potentials less steep than a parabolic potential, such as a triangular potential, the level spacing will actually decrease for increasing $N$. Therefore, our data suggests a confinement somewhere between a parabolic and a hardwall potential, at least for the higher electron numbers further away from depletion, which is consistent with the simulated potentials used for the quantum interference oscillations. However, we note that we are filling electrons only underneath the center gated segment, not along the full  length of the wire, see Fig.~\ref{fig:6um_zb}. 

At around $N_e\sim$ 430 electrons, the even-odd effect seems to be reduced somewhat (size of even diamonds approaching that of odd diamonds, blue and orange markers in \ref{fig:6um_zb}(g)), but then recovers as more electrons are removed. This apparently has to do with  the evolution of the confinement potential over large numbers of electrons, and is not currently understood. It is important to mention here, that the potential is changed relatively strongly as more electrons are depleted from the wire, i.e. the potential under the central gate is pulled up about half way to the chemical potential, see Fig. \ref{fig:fft} (c).

The dominant contribution to the Coulomb peak's magnetic field dependence is a diamagnetic shift, which stems from the confinement along the directions perpendicular to the external magnetic field, and grows quadratically with an applied magnetic field \cite{Duncan2000,Folk2001,Lindemann2002, Potok2003}. This effect is dominated by the strong confinement direction, which in our case is the confinement along $\hat z$ due to the quantum well. Because all the discrete levels of the UW occupy the same lowest subband of the quantum well, we expect them to all experience the same diamagnetic shift. Indeed, this is observed when comparing the diamagnetic shift of a $2|2|2$ and $2|6|2$ configuration, once the different lever arms are taken into account, see Appendix \ref{10mu}. The observed direction of the shift in Coulomb peak position, i.e. whether the peaks curve up or down, depends on the relative diamagnetic shift between the participating regions \cite{Lindemann2002}, in our case the UW and LW where the $\hat z$-confinement is slightly different due to different quantum well widths. 

Additionally, we observe that neighboring traces exhibit an alternating behavior, either approaching or moving away from each other. This is due to the Zeeman-effect giving a spin dependent contribution $E_z = Sg\mu_BB$ to successive spin states, where $S$ is the spin quantum number, $g$ the Landé g-factor, and $\mu_B$ is the Bohr magneton. Assuming the diamagnetic shift to be the same for all resonances, the energy shifts with B of the spin states can be written as,
\begin{align}
E^\pm_{N} &= \pm g\mu_BB/2 + \gamma B^2. 
\end{align}
With this, the diamagnetic shift can be isolated by summing two adjacent resonances. As long as they have alternating spin, the Zeeman terms cancel. We extract the prefactor for the diamagnetic shift to be $\gamma = -15\,\mu$eV/T$^2$, and confirm our assumption of essentially identical diamagnetic shifts, as the traces in Fig. \ref{fig:6um_zb}(e) overlap. By subtracting neighbouring resonances, only the Zeeman splitting between the two states $\Delta E_z = g\mu_B B$ remains. The extracted Zeeman splitting for the first few states is shown in Fig. \ref{fig:6um_zb}(f). The magnitude of the extracted g-factor agrees well with the bulk value of $|g|=0.44$ and is shown in Fig. \ref{fig:6um_zb}(f) as dashed lines. The diamagnetic shift quickly becomes dominant over the Zeeman term with growing B-field and is about an order of magnitude larger at 5\,T. 

When carefully studying the spin filling using the above tool over more than 100 electron transitions, we observe a striking alternation of $S=\pm 1/2$ over all of the 100 observed resonances (see Appendix \ref{full_res} for a larger gate voltage range version of Fig. \ref{fig:6um_zb}(a)). In one case, the even and odd diamonds are of approximately equal size, suggesting a small or vanishing orbital energy splitting. 
Yet the observed Zeeman splitting still continues to strictly follow the alternating spin sequence across this and the neighboring transitions. This behavior of uninterrupted alternating spins is remarkable as we observe this while removing more than a 100 electrons in a 1D system, where electron interactions could be expected to be strong. 

The interplay of orbital and exchange energies determine the spin filling \cite{OregHalperin,BrouwerOregHalperin}. While there may be no magnetized ground states in an ideal, infinite 1D system \cite{LiebMathis, Mermin1966}, real 1D systems such as carbon nanotubes also mostly exhibit  alternating spin filling \cite{Liang2002,Cobden2002,Moriyama,Annabi2024}. In contrast, higher spin states were indeed seen in 2D systems e.g. in semiconductor quantum dots \cite{Duncan2000, Folk2001, Lindemann2002, Potok2003}. Energy degeneracies, either deliberate by choosing a symmetric shape \cite{Tarucha1996} or statistical due to asymmetric ("chaotic") shapes \cite{Patel}, can lead to a type of Hund's rule mechanism via exchange interactions favoring higher spin states \cite{Jiang2003} if electron interactions are sufficiently strong. This happens in form of a tradeoff between kinetic energy of the higher occupied orbital state and an energy gain due to the alignment of the spins \cite{Babic2004}. 

Therefore, we hypothesize that in our 1D system the strict spin sequence originates in the absence of orbital energy degeneracies combined with relatively weak exchange interaction effects. The wire only occupies the transverse ground state and, given its 1D nature, should have a well defined, mostly non-zero longitudinal level spacing, increasing with $\Delta E \propto n$ for hard wall confinement. The interaction strength is reduced $r_s \propto 1/n $ at higher densities, in addition to screening from the metal gate about 500\,nm above the parallel wires. The low density regime would be interesting to investigate. Particularly in the ungated segments, where screening is weaker, though there, the electron number cannot easily be changed. Indeed, in the low density regime ($n\lesssim20\mu\mathrm{m}^{-1}$), an anti-ferromagnetic Wigner crystal is predicted \cite{Matveev}, which at even lower densities crosses over into a ferromagnetic state (at finite magnetic field) \cite{Mueller2005, Qian2008}. 

\section{conclusion}
We have investigated two manifestations of finite size effects in the momentum resolved tunneling spectroscopy of exceptionally long quantum wires with aspect ratio of $\sim$1'000 and up to 18\,$\mu m$ length -- approaching macroscopic length scales. In the continuum regime we observe an intricate set quantum interference patterns. Using a model of non-interacting electrons in 1D, we can explain these features appearing in the tunneling conductance of parallel wires in terms of the Fourier components of the wave functions of the segments in the wires. The striking agreement between experiment and parameter-free simulation showcases coherent quantum interference over the full length of the wires up to 18\,$\mu$m length, demonstrating excellent quality of the wires and the excellent coherence facilitated by the state-of-the-art low noise and low temperature experiment.  

In addition, we observed the discrete nature of the electronic states, clearly revealing the electronic spectrum of up to 10\,$\mu$m long single mode quantum dots. A striking result is the regularity of alternation between spin-up and spin-down states, without the appearance of higher spin ground states. This could possibly be explained by weak exchange and the lack of accidental degeneracies in the energy spectrum of a single mode 1D wire as opposed to 2D quantum dots, where orbital degeneracies can occur regularly.  

These results lead to the question how long such a wire could be made while still exhibiting quantum interference and/or discrete quantum states? We note that length scales of about 20-50\,$\mu$m  are approaching the macroscopic regime, visible by the bare human eye (diameter of a hair). To estimate these limits, we assume typical phase coherence times $\tau_\varphi = 3\,$ns in high mobility GaAs 2DEGs \cite{Huibers1999}. The coherence length $L_\varphi=v_F\tau_\varphi \approx 600\,\mu$m would suggest our observations are not limited by the coherence or the thermal length $L_{T}  = \hbar v_F/k_B T $, which reaches about 150\,$\mu$m at 10\,mK. 

In addition, the  diminishing oscillation periods as the length is increased should not pose a problem, since for a wire length of 100$\mu$m the magnetic field spacing of the full length oscillations will still be $\sim 1\,$mT, which is easily detectable. Yet, all of these lengths by far exceed the transport mean free path which is known to be larger than $10\,\mu$m \cite{Yacoby1996} but probably well below $100\,\mu$m. At some length the wire is not ballistic anymore, which should lead to deviations from the simulation results, as  observed in the 18$\,\mu$m wire for more negative gate voltages. While this has little influence on the visibility of interference, it might introduce additional periods or modify existing ones, related to new lengths appearing due to the presence of scatterers.

Similarly the observation of longitudinal discrete states is only limited by the shrinking of the level spacing as length is increased. A spacing of 1\,$\mu$V, a reasonable limit of detection given present limitations on temperature and noise, will be reached at about 33\,$\mu$m. 

\begin{acknowledgments}
We thank G. Barak for providing access to this sample and we acknowledge S. Martin and M. Steinacher for technical support. We also thank D. Loss and J. Klinovaja for the usefull discussions. This work was supported by the Swiss SNSF (grant no. 179024 and 215757) and the Georg H. Endress Foundation and the NCCR SPIN of the Swiss SNSF, the EU H2020 European Microkelvin Platform EMP (Grant No. 824109) and the UpQuantVal InterReg. 
\end{acknowledgments}
\bibliography{./henoks_refs_merge2_dmz}
\clearpage
\section{Appendix}
\appendix

\section{Methods} \label{Method}
\subsection{Sample fabrication}
A GaAs double quantum well (QW) is grown by molecular beam epitaxy along the [100] direction, with the upper Well being 20\,nm wide and the the lower 30\,nm. The $\delta$-doping is adjusted such that only the upper QW is populated and conductive, with an electron density $n_{2DEG}\approx1\times10^{11}\,$cm$^{-2}$ and mobility $\sigma\approx 3\times10^6\,$cm$^2$/(Vs) while electrons in the lower QW remain localized. 
The sample is cleaved inside the MBE machine, and successively overgrown with GaAs and an additional layer of Si-dopants onto the [110] surface. The electric field from the ionized dopants on the cleaved edge attracts additional electrons, thereby forming 1D modes in both the upper and lower QW, referred to as the upper wire and lower wire. The resulting density in the quantum wires is about $n_w\approx80\,\mu$m$^{-1}$. Further details on the sample fabrication can be found in \cite{Pfeiffer1993,Auslaender2002}.

\subsection{Cryogenic setup}
We performed measurements in a Leiden Minikelvin MNK126-700 TOF $^3$He/$^4$He dilution refrigerator with the mixing chamber temperature reaching as low as 8\,mK. Furthermore, a combination of discrete component RC-filters, and microwave-filters \cite{Scheller2014} were used to achieve thermalization of the electrons and low noise levels. Most of the displayed data was acquired using standard lockin (Signal recovery 7265)  techniques at a few HZ ($\sim 17\,$Hz) using excitations of 2-6$\,\mu$V. Basel Precision Instruments I/V converters and differential amplifiers were used as a preamplifaction stage at room temperature. Dc voltages where recorded with Agilent 34410A digital multimeters.

\section{Simulation}\label{simulation}
\subsection{Theory and Simulation} 
The following section describes the details and results of the single particle, non-interacting simulation, that accurately describes most seen features. 
The UW is modeled as free electrons in 1D confined to length $L$ with the Hamiltonian,
\begin{equation}
H = \frac{\hbar^{2}}{2 m^{*}}\frac{\partial^{2}}{\partial x^{2}}+U_{UW}(x)
\end{equation}
where $m^{*}$ is the GaAs effective electron mass $m^{*}=0.067 m_{e}$ , with the free electron mass $m_{e}$, and $U_{UW}(x)$ is the longitudinal confinement potential due to the gates. Previous works \cite{Tserkovnyak2002,Tserkovnyak2003} used a power-law potential $U(x) = U_0 |2x/L|^\beta $, where the exponent $\beta$ governs the steepness of the confinement potential. Later, a softened potential bump, was used to model the effect of the central gate\cite{Qian2008}.  In this work, we model all gates using an analytic formula given by Davies et al.\cite{Davies1995}, specifically derived for stripe gates on top of a 2DEG in GaAs, providing a smooth gate potential. 

The specific shape of the potential created by the i-th gate is calculated by
\begin{align}\label{eq:davies}
U_i(x) = &\frac{V_G a_{0}\pi}{8 d K(1/\mathrm{cosh}({\pi a /2d}))}\nonumber\\
&\times({1+\mathrm{sinh}^{2}({\pi x /2d})^{2}/\mathrm{cosh}^{2}({\pi a /2d}))^{-1/2}},
\end{align}
where $d$ is the distance from the surface gate to the wire ($\sim$500\,nm), $a$ is the width of the gate (2\,$\mu$m or 6\,$\mu$m), $V_G$ is the applied gate voltage, K is the complete elliptical integral of the first kind, and $a_{0}= 4\pi \epsilon_0\epsilon\hbar^{2}/{m^*e^2}$ is the scaled Bohr radius for a semiconductor with dielectric constant $\epsilon$ ($\sim$12 for Al$_{0.32}$Ga$_{0.68}$As)\cite{Chklovskii1992}. The only free parameters are the distance between the gate and the 2DEG, and the width of the gate.

The UW potential is then just $U_{UW}(x)=\sum_i U_i(x)$. Because Eq. \ref{eq:davies} does not incorporate the screening properties in the devices due to the multiple modes per wire, and the screening of the LW by the UW, a slightly modified approach is used to find the strength of the induced potential.  The potential change due to an applied gate voltage is incorporated thus by using the extracted densities in the gated section for the UW and LW from the large field measurement Fig. \ref{fig:large_scale} (a). The electron density is related to the local potential by $U(x,V_G)=\hbar^{2}\pi^{2}(n_0^2-n(x,V_G)^{2})/8m_{e}$, where $n_0$ is the density at $V_G = 0$ and $n(x,V_G)$ is the gate-voltage dependent density. This establishes a relationship between the voltage on the central gate and the corresponding density under that gate. So instead of inserting a value for the gate-voltage $V_G$ in Eq. \ref{eq:davies}, we rescale the potential such, that the top flat part centrally under the gate is the potential that creates the correct density $n(x,V_G)$ underneath the gate for that gate-voltage.  This means $U_i(x=x_0) = U(x_,V_G)$ where $x_0$ is the center position of the gate. 

The wave functions in the UW are acquired by solving the stationary Schrödinger equation, $H \psi (x)= E\psi(x)$. As the gate voltage and thereby the potential bump is increased, the wave function transforms the most in the gated area. Here, the number of nodes reduces, decreasing the local Fermi wave vector, while it stays almost constant in the ungated region. This is visualized in Fig. \ref{fig:fft}(c), where the wave function closest to the Fermi level, and the potential due to the gates, are shown for different central gate voltages. 

With the wavefunctions at hand, the tunneling matrix element M can now be calculated following Tserkovnyak et al. \cite{Tserkovnyak2002, Tserkovnyak2003} as,
\begin{equation}
M(Q_b)=\int_{0}^{L} \psi(x) e^{-iQ_bx}\varphi(x)dx,
\label{eq: M}
\end{equation}
where $\varphi(x)$ is the LW wavefunction, $\psi(x)$ is the UW wavefunction, and $Q_{b}=eB_yd/\hbar$ is the momentum picked up by a tunneling electron. Since the lower wire is continuous, we obtain the wave function via the WKB-approximation by,
\begin{align}
\varphi(x)= & ({{1} / {2m^*(E_{LW}-U_{LW}(x))}})^{1/4})\cdot\nonumber     \\
& \exp({i/\hbar \int{\sqrt{2m_{e}(E_{LW}-U_{LW}(x'))}dx'}}) ~,
\end{align}
where $U_{LW}(x)$ is the electrostatic potential in the LW due to the gates. This is reasonable because in the LW we stay strictly in the regime where $E_{LW} > U_{LW}(x)$, avoiding the classical turning points $E_{LW} \sim U_{LW}(x)$. We note that it is critical to include the effect of the central gate on the LW density to reproduce the correct period of oscillations for the finite-size effects.

To fulfill energy conservation, the energy of the LW state has to be $E_{LW}= E_{UW}+eV_{SD}+\Delta E_{F}$ with $\Delta E_{F}= E_{F}^{LW}- E_{F}^{UW}$.  The tunneling current $I_{T}$ then is proportional to the sum of squared matrix elements of all states $\psi_{i}$ that are inside the bias voltage window $E_{F} > E_{i}> E_{F}- eV_{SD}$,
\begin{equation}
I_{T}(V_{G},B) \propto \sum^{i}|M_{i}(V_{G},B)|^{2}. 
\end{equation}
To acquire the full range, the wave functions and corresponding eigenenergies in the UW will be calculated for every gate voltage $V_{G}$. Then $|M(V_{G},B)|^{2}$ is calculated for the needed magnetic field range.

The simulations for the three configurations discussed before are shown Fig. \ref{fig:meas_sim}, and display all previously discussed features, such as the correct position of the main resonance, the appearance of the main resonance replicas, and the almost vertical stripes of suppressed interference due to the ungated section. Also, the inner region oscillations are correctly reproduced. Most notably, this is achieved without the need for any tunable parameters. All relevant quantities are either given from the lithographic dimensions, or extracted from the tunneling spectroscopy measurements directly, such as the density of the UW and LW modes. Thus, the agreement, over the three different length configurations, is quite striking, and demonstrates a very thorough understanding of the double wires system.

The previously mentioned resonances due to the presence of higher modes (running diagonally toward the upper right in Fig. 2 (b) in the main text) are not reproduced, because only  the lowest modes were considered.

The fact that the main resonance for the $2|6|2$ and $6|6|6$ configurations (the two arrangements of largest length) appears to be made up of two resonances, which also is not reproduced by the simulation. A possible explanation could be given by the presence of a density variation in the UW underneath the gate. The needed difference in density to reproduce the observed splitting of $\Delta B \sim 100$\,mT  can be estimated as $\Delta n= 2*\Delta B e d / \pi \hbar = 3 $\,$\mu$m$^{-1}$, while the total density without the effect of the gate is $\sim 80$\,m$^{-1}$. A simulation showing this double peak structure, and the assumed potential profile, are shown in the appendix (Fig. \ref{fig:double_resonance}).

\section{Additional data} \label{add_data}
\subsection{Data from a 10 micron wire} \label{10mu}
The same procedure as in the main text was used to perform the analysis of the discrete states in a 10\,$\mu m$ quantum wire with a 6\,$\mu m$ gate ($2|6|2$). The general behavior is quite similar to the $2|2|2$ case discussed in the main text, with a few differences. Due to the smaller orbital and charging energies the data extraction is significantly more challenging. 

\begin{figure*}[tb!]
\centering
\includegraphics[width=\linewidth]{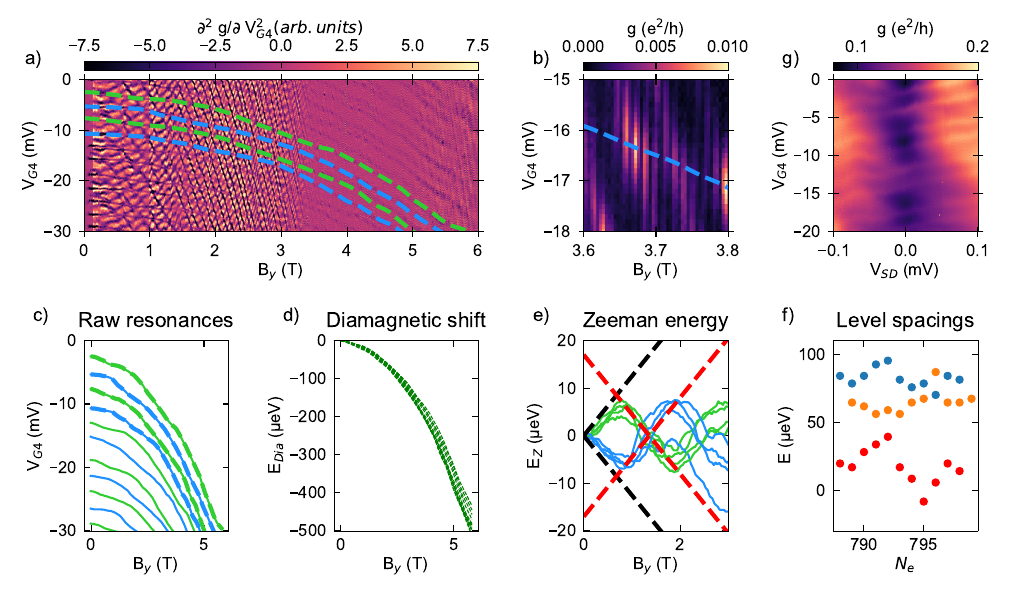}
\caption{ Energy level spectroscopy of a 10\,$\mu$m ($2|6||2$) long wire. (a) conductance map showing the first few Coulomb peaks. For magnetic fields below 3\,T additional peaks from other modes reduce visibility. The second derivative with respect to gate voltage is shown for clarity. Four resonances are marked with alternating green and blue dashed curves. (b) Zoom in into the dashed white region in (a). Dashed curves are the same as in (a). Each Coulomb peak shows periodic modulation of conductance as $B_y$ is varied, with $\Delta B = 14$\,mT corresponding to a length of 9.5\,$\mu m$. (c) Coulomb diamonds diagram taken at $B_y=0.13$\,T of the first few resonances used to extract the lever arm $\alpha=17.8\,\mu$eV/mV. d) The extracted resonances from (a). (e) The extracted diamagnetic shift for the first 11 resonances. (f)  Extracted  Zeeman energy for the first 8 resonances. Dashed black lines correspond to the bulk GaAs g-factor of  $g=-0.44$. Red dashed lines mark the crossover from higher/lower orbitals that are lowered/lifted by the Zeeman energy. (g) Dependence of addition energy $E_{add}$, charging energy $E_{c}$, and single particle levels spacing $E_{orb}=E_{add}-E_c$ on occupation number.}
\label{fig:full_resonances_10} 
\end{figure*}

The first difference that one can notice is the smaller seeming diamagnetic shift. expressed in gate voltages, the shift appears to be only -30\,mV in the 10\,$\mu m$ case at 5\,T compared to -60\,mV in the 6\,$\mu m$ case. This seems to contradict the statement that the diamagnetic shift only depends on the quantum well confinement, which should be the same, regardless of the length of the wire. Upon closer examination, this discrepancy can be resolved when including the leverarm of the gate action. The leverarm, which describes the ratio of voltage change on the gate to potential change on the wire,  can be expressed as the ratio of wire-gate capacitance $C_G$ and the total capacitance the wire sees $C_\Sigma$. While $C_g$ scales linear with  the length of the gate $L_G$, $C_\Sigma$ scale linear with the total length of the wire $L_T$  (The dominant contribution to the total capacitance stems from the wire-wire capacitance between UW and LW). The two length in both cases are ($L_T,L_G$) (6\,$\mu m$, 2\,$\mu m$) and (10\,$\mu m$, 6\,$\mu m$), meaning that the ratio of these lengths, and therefore the leverarm in the $2|6|2$ configuration is about a factor of 1.8 larger than in the $2|2|2$ case. This agrees well with the leverarm of 17.8\,$\mu$eV/mV extracted from the Coulomb diamonds extracted in Fig. \ref{fig:full_resonances_10} (c) when compared to the $2|2|2$ leverarm of 9.6\,$\mu$eV/mV. 

\subsection{6 micron wire as a function of density difference}

\begin{figure*}[tb!]
\centering
\includegraphics[width=1\linewidth]{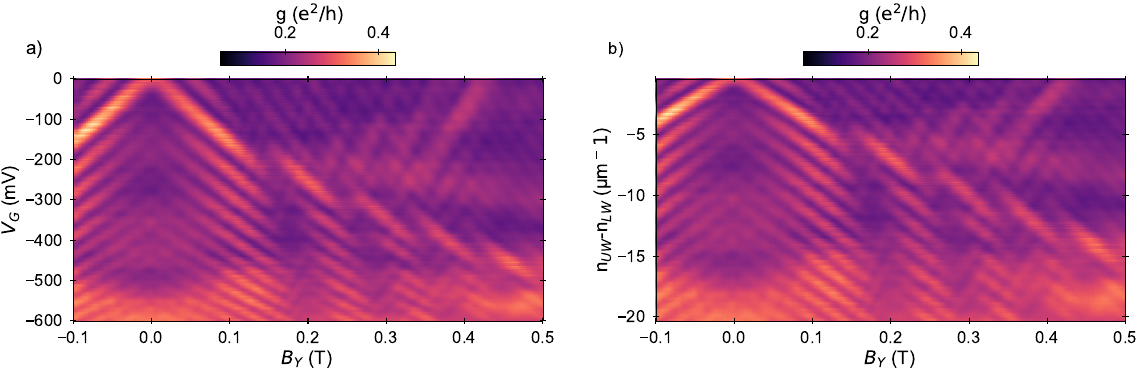}
\caption{ Rescaled tunneling conductance map for the $2|2|2$ configuration. Left: Data as measured and presented in the main text. Right: The y-axis has been rescaled to the difference of the UW and LW densities, removing the curvature of the resonances.   }
\label{fig:rescaled_resonances} 
\end{figure*}

\subsection{Full map of tunneling resonances} \label{full_res}
\begin{figure*}[htb!]
\centering
\includegraphics[width=1\linewidth]{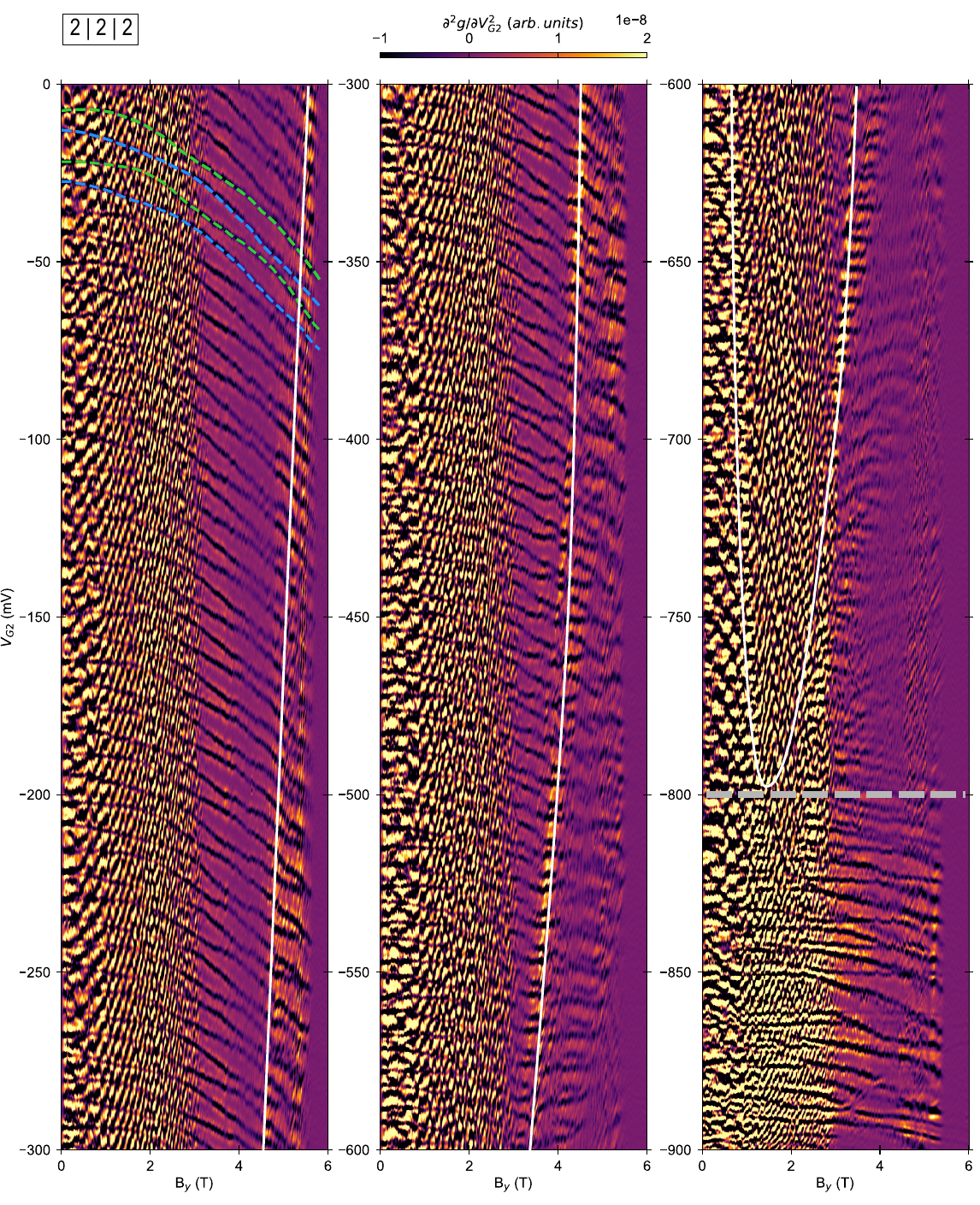}
\caption{Full gate voltage range of the Coulomb resonance map shown in Fig. \ref{fig:6um_zb} (a) for the $2|2|2$ configuration, with green and blue dashed curves marking the same resonances. The white solid curve marks the position of the main resonance. In the first two panels, only the counter propagating transition is indicated. At voltages below -800\,mV a localization transition occurs, which is indicated by a thick dashed gray line.  }
\label{fig:full_resonances} 
\end{figure*}

In Fig. \ref{fig:full_resonances} we show the full evolution of the Coulomb resonances starting from the highest density at zero gate voltage,  down to depletion underneath gate $G_2$. The resonances marked with blue and green dashed curves are the same as shown in Fig. \ref{fig:6um_zb}(a). Depletion of the UW underneath gate $G_2$ occurs at around -800\,mV, as can be seen also in Fig. \ref{fig:large_scale}(a), where below this gate voltage  horizontal streaks appear in the tunnelling spectroscopy measurement. These features were observed \cite{Auslaender2005, Steinberg2006} and studied before \cite{Fiete2005, Mueller2005, Qian2008,Gueclue2009} and are related to a localization transition at very low densities.

In Fig. \ref{fig:full_resonances} the main resonance for momentum matching is indicated by a white curve in all three panel, but for the first two panels only the counter propagating transitions is shown, because here the co-propagating transition would be very close to the figure axis, due to the large aspect  ratio of the figure.

One feature that can be observed is a reduced strength of the diamagnetic shift, as gate voltage becomes more negative. This can be caused by a shift of the wave function position in the quantum well as density is changed, or due to the additional potential due to the field \cite{Lindemann2002}.  The assumption used to extract the Zeeman splitting, that the diamagnetic shift for adjacent resonances is similar, is still valid since the change only becomes notable when looking at many resonances. 

Close to the counter propagating transition at intermediate to large gate voltages, the Coulomb peaks behave unusual. The previously smooth resonances start displaying kinks and jumps that are not seen at less negative gate voltages. This behavior is not understood.

\subsection{Asymmetric ungated sections} \label{asym}
\begin{figure*}[htb!]
\centering
\includegraphics[width=0.8\linewidth]{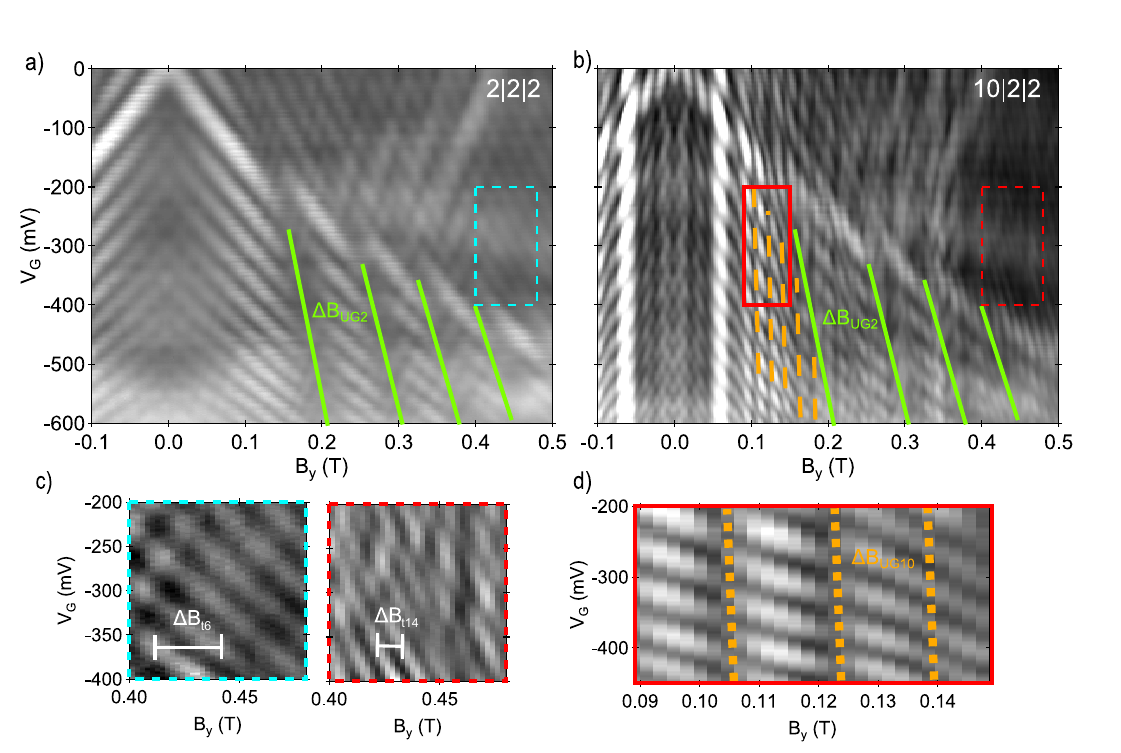}
\caption{Comparison between symmetric $2|2|2$ (a) and asymmetric $10|2|2$ configurations. c) Zoom in on the dashed boxes in (a) and (b), showing the inner region oscillations. (d) Zoom in on the solid rectangle in (b), showing the additional pattern appearing due to the two lengths of ungated sections.}
\label{fig:asym} 
\end{figure*}

Previously we only presented the finite size effects as they appear in wires with two symmetric ungated sections. Here we will discuss the effect of having two ungated sections of different lengths. This will predominantly lead to the appearance of two different ungated oscillation periods. Though not surprising, it is important to check that the expected behaviour of overlaid oscillation patterns is consistent with what is observed in measurements and simulations.

We will compare the case of $2|2|2$ with $2|2|10$. From the previous analysis, we would expect the gated oscillations to remain the same, while we should still observe 2\,$\mu$m  ungated oscillations accompanied by much faster 10\,$\mu$m  oscillations. In addition, the full-length oscillations should now correspond to a length of 14\,$\mu$m instead of 6\,$\mu$m , which would increase the frequency.

Figure \ref{fig:asym} (a) and (b) show the tunnelling conductance for the $2|2|2$ and $10|2|2$ configuration respectively. In addition to the modulation stemming from the 2\,$\mu$m ungated section which is marked by green solid lines in both graphs, faster oscillations stemming from the 10\,$\mu$m ungated section are marked by orange dashed lines in Fig. \ref{fig:asym} (b). A zoom-in into the red solid box is shown in Fig. \ref{fig:asym} (d). The extracted Periods are: $\Delta B_{UG2}=94$\,mT and $\Delta B_{UG10}=14$\,mT.  Under the assumption that the discrepancy between the lithographic and effective length of a section is more pronounced for shorter sections, we assume that the 10\,$\mu m$ section is effectively 10\,$\mu m$ long, while for the 2\,$\mu m$ section we use the extracted value from \ref{tab:lengths}, which is $L_{UG2}=1.5\,\mu m$. Then the ratio of length and the ratios of oscillation periods both yield $L_{UG2}/L_{UG10} \sim \Delta B_{UG10}/\Delta B_{UG2} \sim 6.6$. This agreement between the periods and the estimated length of the sections confirms that the dominant effect of sections of different lengths is the overlaying of features stemming from the respective regions, without influencing each other.

Another Period we can investigate is the inner region oscillations, related to the full length of the wire. These are presented in Fig \ref{fig:asym}(c) as magnifications of the dashed boxes of Fig. \ref{fig:asym} (a) and (b). The total lithographic lengths of the wires are $L_{t6} = 6\,\mu m$  and $L_{t} = 14\,\mu m$ respectively.  Again we find that the ratios agree quite well, $L_{t6}/L_{t14} =2.33 $ which is quite close to $\Delta B_{t14}/\Delta B_{t6} =2.4 $

\subsection{Dual gated wire}
To complete our study of the finite size effects, we also show the effect of using a second gate on the same wire. In this way, we create two regions of reduced density. The question that arises is whether the gate-induced oscillations of one gate can be compensated by the use of a second gate. 

\begin{figure*}[htb!]
\centering
\includegraphics[width=12cm]{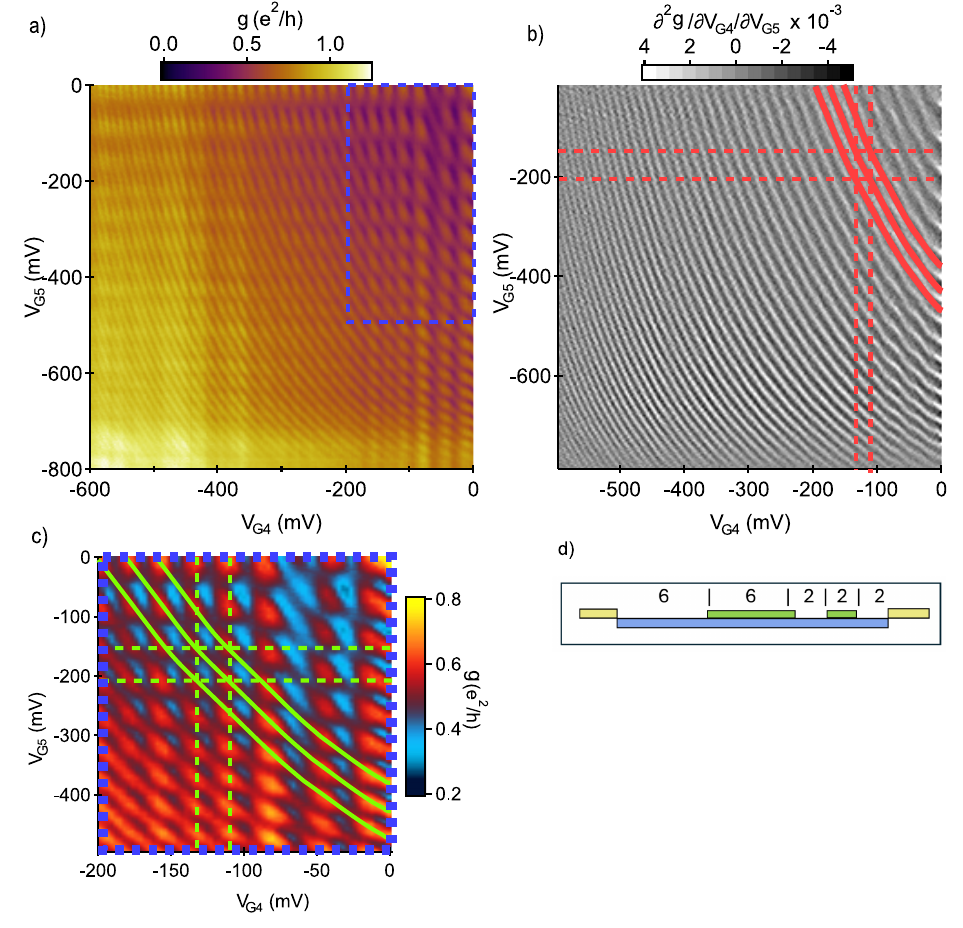}
\caption{Two-gate finite size effects: a) $G(V_{G4},V_{G5})$ , b) second derivative of a), c) Zoom into the blue dashed region in a) }
\label{fig:twogate}
\end{figure*}

We use gates 2 and 7 as barrier gates, which create a 18\,$\mu$m wire between them. This wire section contains two center gates of length 2\,$\mu$m (g5) and 6\,$\mu$m (g4), such that we have $ 6|6(G4)|2|2(G5)|2$. By applying an in-plane magnetic field $B_y = -50mT$ we move away from the symmetric point of $G(B_y,V_G)$ at $B_y=0$. Figure \ref{fig:twogate} (a) shows the measurement of $G(V_{G4},V_{G5},{B_y=-50})$\,mT. 

\subsection{Temperature dependence of Coulomb blockade}
\begin{figure*}[htb!]
\centering
\includegraphics[width=13cm]{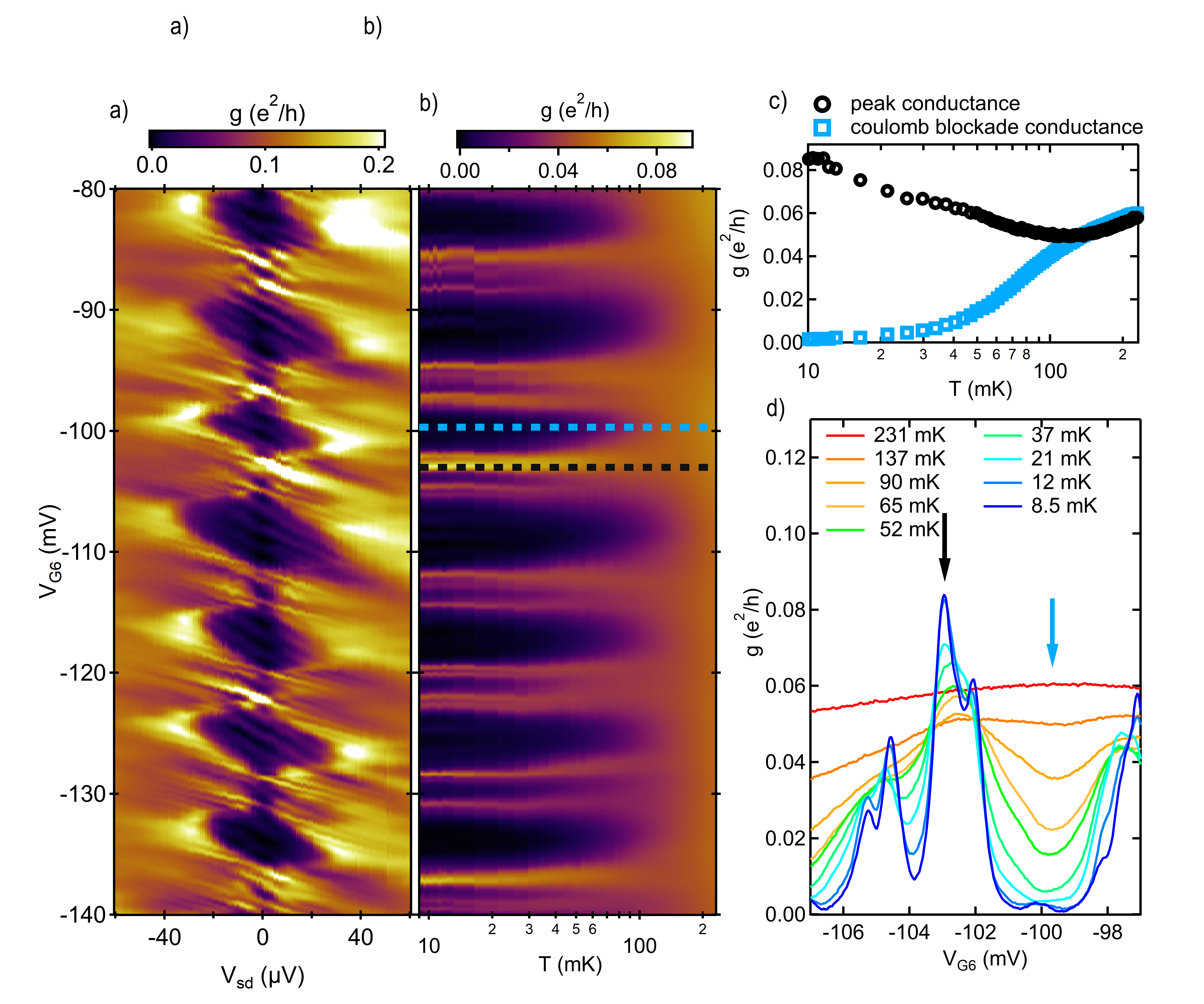}
\caption{ Temperature dependence of Coulomb peaks for the $2|6|2$ configuration. (a) Coulomb diamonds. (b) Conductance map for the same Coulomb peaks as in (a) as a function of temperature at zero bias. (c) Line cuts along the blue and black dashed line in (b). The blue markers show the temperature dependence inside a Coulomb blockade valley, while the black markers show the Coulomb peak conductance. (d) Vertical line cuts showing the temperature evolution across multiple Coulomb peaks.}
\label{fig:temp_dependence}
\end{figure*}

In this section, we show the temperature dependence of Coulomb peaks at the lowest accessible temperatures. During this cooldown, the mixing chamber base temperature reached 8.5\,mK. This was the temperature at which all the data presented in the main paper were recorded. Figure \ref{fig:temp_dependence}(a) shows the conductance map as a function of gate voltage and temperature, measured at zero bias. The configuration for this measurement is a $2|6|2$ configuration, with a lithographic length of 10\,$\mu$m. The even-odd spacing is also visible here, with their charging energy being substantially smaller than the energy level spacing. Figure \ref{fig:temp_dependence}(b) shows two horizontal line cuts along the blue and black dashed lines in Fig. \ref{fig:temp_dependence}(a). The blue data points in (b) show the temperature dependence of the conductance inside a Coulomb blockade valley. The conductance increases monotonically as temperature is increased.

On the contrary the black markers show the temperature evolution of conductance on a Coulomb peak. At the lowest temperatures, the peak height reduces as the temperature is lowered, and reaches a minimum around 130\,mK, which corresponds to an energy of 10.3\,$\mu$eV. For higher temperatures, the conductance on the peak increases again, and the conductance of the peak and the valley approach each other.

This behaviour of the non-monotonic peak height evolution was predicted by Furusaki and Nagaosa  \cite{Furusaki1993} for a finite length Luttinger liquid. The authors concluded that the minimum should appear at a temperature $T \sim \Delta E $,  with $\Delta E$ being the level spacing of the quantum wire. This type of scaling should only be observed above a certain temperature, given related to the thermal coherence length $T_L = \hbar v_F / k_B T$ \cite{Kane1992a}. For  a 10\,$\mu m$ wire  $T_L = 160$\,mK. At temperatures smaller (larger) than $T_L$, the thermal coherence length is larger (smaller) than the length of the wire. 

A similar study on the CEO-wires was performed by Auslaender et al. \cite{Auslaender2000} for the Coulomb peaks close to the depletion of the wire (from $\sim$-800\,mV in Fig. \ref{fig:full_resonances}), where they investigated the peak width of the coulomb peaks as a function of temperature. 

\subsection{Doubling of the main resonance} \label{doubling}
\begin{figure}[htb!]
\centering
\includegraphics[width=6cm]{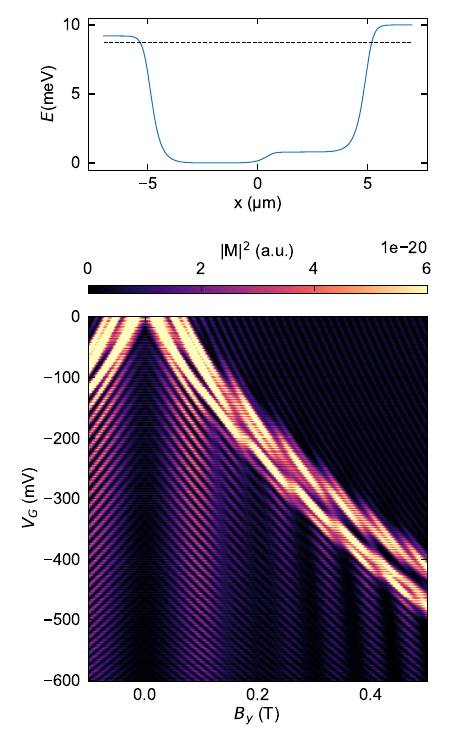}
\caption{ Simulation of the doubled main resonance for the $2|6|2$ configuration. (a) shows the assumed potential profile without the activation of the 6\,$\mu$m central gate. An increased potential, corresponding to a decreased density of 76\,$\mu$m$^{-1}$ was used in the right half (positive $x$-values) compared to 80\,$\mu$m$^{-1}$ on the left side (negative $x$-values). (b) Calculated tunneling matrix element for the potential shown in (a), as the central gate voltage and spectroscopy field are varied.  }
\label{fig:double_resonance}
\end{figure}

For the two long configurations, a doubling of the main resonance was observed. We hypothesize that the density might vary along the long wires by a small amount. This would lead to multiple wave vectors, at which the matching can occur. To give an example, we model the $2|6|2$ configuration, but reduce the density on one half of the wire by 3\,$\mu$m$^{-1}$. To do this, the potential has to be raised by $\sim 0.6$\,meV calculated with $\Delta E = hbar^2\pi^2(n_0^2-n^2)/8m^*$, with $n_0$ the original density, and $n$ the reduced density. This setup is show in Fig. \ref{fig:double_resonance} (a). We then run the simulation as before, varying the central gate voltage, and calculating the tunnel matrix element for each $V_g$ and $B_y$. The result is shown in Fig. \ref{fig:double_resonance} (b), showing strong resemblance with the measured data shown in Fig. \ref{fig:meas_sim} (a). 

\end{document}